\documentclass[onecolumn,showkeys,preprintnumbers,aps,a4paper,amssymb,prd,superscriptaddress,nofootinbib]{revtex4-2}
\usepackage{comment}
\usepackage{graphicx}
\usepackage{epsf}
\usepackage{bm}
\usepackage{amsmath}
\usepackage{amsfonts}
\usepackage{amssymb}
\usepackage{epstopdf}
\usepackage{color}
\usepackage[dvipsnames]{xcolor}
\usepackage{verbatim}
\usepackage{multirow}
\usepackage{soul}
\usepackage{physics}
\usepackage{bm}

\usepackage[width=0.00cm, height=0.00cm, left=1.50cm, right=1.50cm, top=2.00cm, bottom=2.00cm]{geometry}
\usepackage{microtype}
\usepackage{lmodern}

\usepackage[colorlinks = true,
            linkcolor = teal,
            urlcolor  = teal,
            citecolor = teal,
            anchorcolor = blue]{hyperref}
\usepackage[capitalize]{cleveref}
\usepackage[normalem]{ulem}
\usepackage{enumitem}
\usepackage{booktabs}

\usepackage{lipsum}

\makeatletter\let\expandableinput\@@input\makeatother

\begin{document}



\begin{center}
		\vspace{0.4cm} {\large{\bf Empirical Validation: Investigating the $\Lambda_s $CDM Model with new DESI BAO Observations
}} \\
		\vspace{0.4cm}
		\normalsize{ Manish Yadav$^1$ Archana Dixit$^2$ Anirudh Pradhan $^3$ M S Barak$^4$ }\\
		\vspace{5mm}
		\normalsize{$^{1,4}$ Department of Mathematics, Indira Gandhi University, Meerpur, Haryana 122502, India\\
		\normalsize{$^{2 }$ Department of Mathematics, Gurugram University Gurugram, Harayana, India}\\ 
		Mathura, U.P., India-281406.} \\

        	\normalsize{$^{3 }$ Centre for Cosmology, Astrophysics and Space Science (CCASS), GLA University, Mathura, Uttar Pradesh, India}\\ 
		\vspace{2mm}
			$^1$Email address: manish.math.rs@igu.ac.in\\
		    $^2$Email address: archana.ibs.maths@gmail.com\\
                $^3$Email address: pradhan.anirudh@gmail.com\\
             $^4$Email address: ms$_{-}$barak@igu.ac.in\\
\end{center}

 
\maketitle
{\noindent {\bf Abstract.}
The $\Lambda$CDM model has long served as the cornerstone of modern cosmology, offering an elegant and successful framework for interpreting a wide range of cosmological observations. However, the rise of high-precision datasets has revealed statistically significant tensions, most notably the Hubble tension and the $S_8$ discrepancy, which challenge the completeness of this standard model. In this context, we explore the $\Lambda_{\rm s}$CDM model—an extension of $\Lambda$CDM featuring a single additional parameter, $z_\dagger$, corresponding to a sign-switching cosmological constant. This minimal modification aims to alleviate key observational tensions without compromising the model’s overall coherence. Recent findings present in the literature indicate that the $\Lambda_{\rm s}$CDM model not only provides a better fit to Lyman-$\alpha$ forest data for $z_\dagger < 2.3$, but also accommodates both the SH0ES measurement of $H_0$ and the angular diameter distance to the last scattering surface when 2D BAO data are included. We present a comprehensive analysis combining the full Planck 2018 CMB data, the Pantheon Type Ia Supernovae sample, and the recently released Baryon Acoustic Oscillation (BAO) measurements from the Dark Energy Spectroscopic Instrument (DESI). Our finding reveal that the Preliminary DESI results, a possible $3.9\sigma$ deviation from $\Lambda$CDM expectations, reinforce the importance of exploring such dynamic dark energy frameworks. In sum, our study underscores the potential of $\Lambda_{\rm s}$CDM to reconcile multiple cosmological tensions and sheds light on the role of upcoming high-precision observations in reshaping our understanding of the universe's expansion history and the nature of dark energy.


\section{Introduction}

For a long time, the Standard Lambda Cold Dark Matter ($\Lambda$CDM) model has been the backbone of contemporary cosmology, effectively explaining a wide range of cosmological observations, such as cosmic microwave background (CMB) \cite{ref1,ref1a},  the late-time  accelerated expansion~\cite{ref2,ref3} and large-scale structural development of the universe, making it the prevailing paradigm in cosmology. However, despite its achievements, it grapples with several significant tensions~\cite{ref4,ref5,ref6}. Two of the most prominent tensions in contemporary cosmology that raise questions on  completeness of the standard $\Lambda$CDM model are the Hubble tension and the $S_8$ tension. A notable discrepancy known as the Hubble tension resulted from varying results obtained for the Hubble constant, $H_0$, via different observational techniques. This tension represents a major inconsistency between the locally measured $H_0$ value from the SH0ES collaboration, $H_0=73.04\pm1.04~{\rm km\, s^{-1}\, Mpc^{-1}}$~\cite{ref7}, using Cepheid-calibrated Type Ia supernovae, and the early-universe estimate from the Planck collaboration, $H_0=67.36\pm0.54~{\rm km\, s^{-1}\, Mpc^{-1}}$~\cite{ref1}, which is based on CMB observations within the $\Lambda$CDM  model, there exists a statistically significant discrepancy of 5.0$\sigma$. Recent low-redshift observations have indicated a comparatively higher value of the Hubble constant ($H_0$), creating a notable tension with the estimate derived from Planck-CMB data. For instance, the Megamaser Cosmology Project~\cite{ref8} has reported a value of $H_0 = 73.9 \pm 3.0\, \mathrm{km\, s^{-1}\, Mpc^{-1}}$, while measurements based on the Surface Brightness Fluctuation method~\cite{ref9} yield $H_0 = 73.3 \pm 2.4\, \mathrm{km\, s^{-1}\, Mpc^{-1}}$. In contrast, the Planck satellite's analysis of the CMB suggests a lower value of $H_0$, which aligns well with constraints obtained from Baryon Acoustic Oscillations (BAO) and Big Bang Nucleosynthesis (BBN), as well as supporting results from other CMB-based experiments including ACT-Pol DR4~\cite{ref10}, ACT-Pol DR6~\cite{ref11}, and SPT-3G~\cite{ref12}. This persistent disagreement suggests a potential gap in our understanding of cosmic expansion. In addition to the Hubble tension, the $S_8$ tension highlights the fundamental challenges of modern cosmology, by showing inconsistencies in the amplitude of matter fluctuations. Large scale structure (LSS) such as weak gravitational lensing predicts a lower value of amplitude of matter fluctuations under $\Lambda$CDM model, viz., $S_8 = 0.759^{+0.024}_{-0.021}$ (KiDS-1000)~\cite{ref13} than predicted by the CMB data, $S_8 = 0.830\pm0.016$ \cite{ref1} within $\Lambda$CDM. This discrepancy raises questions on the key components of the standard model including dark matter and the mechanisms of cosmic structure growth. These ongoing challenges have motivated researchers to seek alternative explanations, whether through innovative physics or by refining data for possible systematic flaws.\\

The $\Lambda_{\rm s}$CDM model is one of the most minimal deviations from $\Lambda$CDM, with one additional free parameter, $z_\dagger$, and sign switching cosmological constant, $\Lambda_{\rm s}$. This model relaxes significant promise in addressing several persistent tensions within cosmological observations,  precisely the disagreements in the $H_0$ and $S_8$ as well as the so-called $M_B$ tension associated with expansion rate, cosmic structure formation, and matter distribution, respectively. However, the Planck data alone cannot provide more precise measurements of cosmological parameters, and additional parameters in the extended model remain unconstrained due to significant degeneracy with existing parameters. The $\Lambda_{s}$CDM model is theoretically motivated by the  dark energy  framework proposed in Akarsu et al. \cite{ref13a}, where the cosmological constant $\Lambda$ is allowed to have a minimal dynamical deviation through a null inertial mass density evolving as $\rho_{\text{inert}} \propto$  $\rho_{\Lambda}$  with $\Lambda< 0$. For large negative $\Lambda$, this formulation naturally leads to a smooth, spontaneous sign switch in $\Lambda$-from negative in the past to positive at late times. This behavior is further interpreted as a possible transition of the Universe from anti-de Sitter (Ads) to de Sitter (dS) vacua around redshift $z_{\dagger} \equiv 2$, an idea that resonates with theoretical expectations from string theory and vacuum landscape scenarios. The $\Lambda_{s}$CDM model has been systematically developed in subsequent works  in \cite{ref13a,ref18}, and shown to offer a significantly better fit to observational data compared to $\Lambda$CDM. It simultaneously addresses multiple long-standing cosmological tensions—including those related to $H_{0}, S_{8}, MB$ , and Ly$\alpha$-using combined datasets from Planck CMB, BAO, Supernovae, and cosmic chronometers. Therefore, the model is both physically motivated and observationally supported.

Therefore, it is essential to combine cosmological probes, such as Baryon Acoustic Oscillations (BAO) and supernova (SN) observations, with Planck data. A key success of the $\Lambda_{\rm s}$CDM framework is its ability to provide a good fit to Lyman-alpha forest observational data include with other datasets for  $ z_\dagger < 2.3$, thereby enabling a cohesive resolution to these ongoing tensions in cosmology. Significantly, the $\Lambda_{\rm s}$CDM model capacity to reconcile these tensions is substantially enhanced by the systematic incorporation of angular 2D BAO data. Meanwhile, 3D BAO measurements, which depend on $\Lambda$CDM assumptions to produce distance metrics, 2D BAO data offers a less biased and more model-independent perspective.  Indeed, using 2D BAO data can be a simultaneous accommodation of the SH0ES $ H_0 $ measurements and the angular diameter distance to the last scattering surface, though it also necessitates an effective negative energy density for redshift $(z > 2)$. This interaction showcases the $\Lambda_{\rm s}$CDM model’s ability not only to resolve current cosmological tensions but also to open new avenues for understanding the fundamental nature of dark energy and the cosmic expansion rate, emphasizing the need for a diverse observational approach in modern cosmology and some studies are presented in the literature about$\Lambda_{\rm s}$CDM model. \cite{ref13a,ref14,ref15,ref16, ref17,ref18}.\\

The DESI is currently conducting  a significant Stage IV survey \cite{ref19,ref20} aimed at refining cosmology constraints by meticulously analyzing the clustering patterns of galaxies, quasars, and the Lyman-$\alpha$ forest. DESI uses a spectroscopic sample size 10 times bigger than previous SDSS surveys to survey $14,200$ square degrees over a five-year period in the redshift region of $0.1$ to $4.2$. The aim of DESI is to precisely constrain the expansion history of the universe and the formation of large-scale structures for cosmological studies. The newly identified BAO \cite{ref21} feature has been verified at a few percent level by early DESI data, suggesting that the survey is on course to meet its main scientific objectives. Notably, the amplitude of primordial fluctuations, neutrino mass, spatial curvature, matter density, and the equation of state of dark energy will be all tightly constrained by DESI \cite{ref22}. Furthermore, it will rigorously test modifications to dark energy components that have been proposed to explain the accelerated expansion of the universe \cite{ref23,ref24,ref25}.\\

Recently, the preliminary data released by the Dark Energy Spectroscopic Instrument (DESI) collaboration suggests a possible 3.9$\sigma$ tension with the $\Lambda$CDM model based on preliminary findings and the dark energy evolution over the cosmic time-frame \cite{ref26,ref26a,ref26b}. Such findings point toward a potential breakdown of the cosmological constant framework, mainly when included with the Planck CMB priors and the DESI 5 Year SN data \cite{ref26}.Several studies have examined this issue, with most suggesting  that the two BAO data points at z=0.51 and z=0.71 may be responsible for the observed result, though some also highlight a potential bias from the selection of dark energy parameter priors \cite{ref27,ref28,ref29,ref30,ref31,ref32,ref33,ref34,ref35,ref36,ref37,ref38,ref39,ref40,ref41,ref42,ref43,ref44,ref45,ref46,ref47}. Numerous studies have investigated how dynamical dark energy models can encompass complex physical phenomena, such as negative dark energy densities at high redshifts and phantom crossings which could serve as pathways to alleviating the $S_8$ and $H_0$ tensions \cite{ref48,ref49,ref50}. Meanwhile, we cannot confidently claim that the DESI results support one of the dynamic dark energy or cosmological constants. This uncertainty persists because the literature includes recent studies that strongly support dynamic dark energy (usually the $w_0w_a$CDM model), while the wCDM model strongly favors the cosmological constant in refs.\cite{ref51,ref52,ref53,ref54,ref55,ref56,ref57,ref58}\\

 In this paper, we primarily focus on examining $\Lambda_{\rm s}$CDM model through observational analysis, using the newly released DESI BAO data, the Pantheon SNIa sample, and the full Planck 2018 data. We investigate how the free parameter ($z_\dagger$) of $\Lambda_{\rm s}$CDM affects (or does not) other cosmological parameters in given datasets. The organization of the paper is as follows:  we provide an introduction of the $\Lambda_{\rm s}$CDM model and describe the observational datasets and methodology employed in our analysis in Section I. The model, datasets and methodology are described in Section II. Then, Section III presents the results of the analysis and discusses the key findings. Lastly, Section IV offers a final summary and conclusion. 


\section{MODEL, DATASETS AND METHODOLOGY}
\label{sec:datasets}

We describe the $\Lambda_{\rm s}$CDM model, a recently proposed and promising variant of standard cosmology. The $\Lambda_{\rm s}$CDM model is an extension of the standard $\Lambda$CDM cosmology, inspired by the recent conjecture observed in the graduated dark energy (gDE) framework. The conjecture reveals that the universe underwent a smooth shift from anti-de Sitter (AdS) vacua to de Sitter (dS) vacua, which is characterized by a sign switching cosmological constant ($\Lambda_{\rm s}$) that changes sign from $-ve$ to $+ve$ at a redshift $z \sim 2$ \cite{ref13a}. In this model, the cosmological constant ($\Lambda$) of the $\Lambda$CDM model was replaced by $\Lambda_{\rm s}$. \\
The sign switching nature of $\Lambda_{\rm s}$ is mathematically expressed as:

\begin{equation}
    \Lambda\quad\rightarrow\quad\Lambda_{\rm s}\equiv\Lambda_{\rm s0}\,{\rm sgn}
    (z_\dagger-z),
\end{equation}
 where $\Lambda_{\rm s0} > 0$ represents the present value of $\Lambda_{\rm s}$  and ${\rm sgn(x)}$ denotes the signum function. The evolution of Hubble parameter of the $\Lambda_{\rm s}$CDM model is governed by the modified Friedmann equation as:  \\

\begin{equation} 
\frac{H^2(z)}{H_0^2}  =\Omega_{\rm r0}\,(1+z
)^{4}+\Omega_{\rm m0}\,(1+z)^{3}+ \Omega_{\Lambda _{\rm s0}}\,{\rm sgn}(z_\dagger - z),  
\end{equation} 
 where, $\Omega_{r0}$, $\Omega_{m0}$,  and $\Omega_{\Lambda_{s}0}$ denote the present density parameters of radiation, matter,  and dark energy, respectively. These parameters satisfy the equation 
  $ \Omega_{r0}+ \Omega_{m0}+  \Omega_{\Lambda_{s}0} = 1$. Our parameter space consists seven baseline parameters of the  $\Lambda_s$CDM model given by;
 
  $$ \mathcal{P}_{\Lambda_s\text{CDM}}
 = \{\omega_b, \omega_c, \theta_s, A_s, n_s, \tau_{\text{reio}}, z_\dagger\}.$$
 
 Here, the baryon energy density $\omega_{\rm b} $, the cold dark matter energy density $\omega_{\rm cdm} $, the angular size of the sound horizon at recombination $\theta_{\rm s}$, the amplitude of the primordial scalar perturbation $\log(10^{10}A_{\rm s})$, the scalar spectral index $n_{\rm s}$, and the optical depth $\tau_{\rm reio}$. Additionally, we consider the redshift $z_\dagger$ at which the sign-switching of $\Lambda$ occurs.  We use flat priors for all parameters in our statistical analyses: $\omega_{\rm b}\in[0.018,0.024]$, $\omega_{\rm c}\in[0.10,0.14]$, $100\,\theta_{\rm s}\in[1.03,1.05]$, $\ln(10^{10}A_{\rm s})\in[3.0,3.18]$, $n_{\rm s}\in[0.9,1.1]$, $\tau_{\rm reio}\in[0.04,0.125]$, and $z_\dagger\in[1,3]$.

The datasets  and methodology used are as follows:
\begin{itemize}
\item \textbf{Planck 2018 (Pk18)}: The Cosmic Microwave Background (CMB) dataset from the Pk18 legacy release is a comprehensive dataset, widely recognized for its precision and accuracy. We use CMB temperature anisotropy and polarization power spectra measurements,  their cross-spectra, and lensing power spectrum~\cite{ref1,ref59}, viz., the high-$\ell$ \texttt{Plik} likelihood for TT ($30 \leq \ell \leq 2508$) as well as TE and EE ($30 \leq \ell \leq 1996$), the low-$\ell$ TT-only likelihood ($2 \leq \ell \leq 29$) based on the \texttt{Commander} component-separation algorithm in pixel space, the low-$\ell$ EE-only likelihood ($2 \leq \ell \leq 29$) using the \texttt{SimAll} method, and measurements of the CMB lensing. 

\item \textbf{DESI Baryon Acoustic Oscillations (DESI BAO)}: 
The DESI BAO data encompass five distinct tracers, such as bright galaxy samples (BGS), luminous red galaxies (LRG), emission line galaxies (ELG), quarks (QSO), and the Ly$\alpha$ forest, at seven different redshift points from the closed interval $[0.1,4.2]$ \cite{ref60,ref61}. These tracers are utilized to calculate $D_{M}(z)/r_{\rm d}$ , $D_{H}(z)/r_{\rm d}$ , and $D_{V}(z)/r_{\rm d}$, which representing the transverse comoving distance, Hubble horizon, and the angle-averaged distance respectively, defined as, 
\begin{equation}\label{fieldeqn}  
D_H(z) = \frac{c}{H(z)},\\
\end{equation}
\begin{equation}
D_{M}(z)=  \int_0^z \text{d}z' {c \over H(z')}.\\
\end{equation}
\begin{equation}
D_V(z) \equiv \left[z D^2_M(z) D_H(z)\right]^{1/3} .\\
\end{equation}

Here $r_{\rm d}$ represents the sound horizon at the drag redshift, and the measurements from the DESI BAO data are listed in Table \ref{tab1}, as outlined in Ref.\cite{ref26}.

\begin{table*}[hbt!]

\caption{\rm The $12$ BAO observations from the Dark Energy Spectroscopic Instrument utilized in our analysis.  }
\centering
\renewcommand{\arraystretch}{1.3} 
\setlength{\tabcolsep}{10pt} 
\begin{tabular}{l|c|c|c|c}
\hline
\hline
\textbf{Tracer} & $\bm{\,\,z_{\rm eff}\,\,}$  &  $\bm{\,\,D_{\rm V}(z)/r_{\rm d}\,\,}$ & $\bm{\,\,D_{\rm M}(z)/r_{\rm d}\,\,}$ & $\bm{\,\,D_{\rm H}(z)/r_{\rm d}\,\,}$  \\
\hline
\hline
\textbf{BGS} & $0.30$ & $7.93 \pm 0.15$ & --- & --- \\

\textbf{LRG} & $0.51$ & --- & $13.62 \pm 0.25$ & $20.98 \pm 0.61$ \\

\textbf{LRG} & $0.71$ & --- & $16.85 \pm 0.32$ & $20.08 \pm 0.60$ \\

 \textbf{LRG + ELG} & $0.93$ & --- & $21.71 \pm 0.28$ & $17.88 \pm 0.35$ \\

\textbf{ELG} & $1.32$ & --- & $27.79\pm0.69$ &$13.82\pm0.42$\\

\textbf{QSO} & $1.49$ & $26.07\pm0.67$ & -- & --- \\

\textbf{Lya QSO} & $2.33$ & --- &$39.71\pm0.94$ & $8.52 \pm 0.17$ \\
\hline
\hline
\end{tabular} 
\label{tab1}
\end{table*}

\item \textbf{Pantheon Plus and SH0ES (PP\&SH0ES)}: The uniform intrinsic brightness of Type Ia supernovae makes them valuable as standard candles. These supernovae provide crucial measurements of distance moduli, which in turn constrain the uncalibrated luminosity distance and can be written as

\begin{equation}
\label{eq:dl}
d_L(z) = (1+z)\int_0^{z}\frac{dz^\prime}{H(z^\prime)}.
\end{equation}

In this study, we utilized the Pantheon Plus SH0ES compilation sample of Type Ia supernova data from references \cite{ref62}. This dataset calibrates the Type Ia supernova magnitude using additional cepheid hot distances \cite{ref63}.

\end{itemize}

 To constrain the $\Lambda_s$CDM model parameters, we perform Markov Chain Monte Carlo (MCMC) analyses using a modified version of the publicly available \texttt{CLASS}$+$\texttt{MontePython} code~\cite{ ref64,ref65}. We have employed the ${R-1 < 0.01}$ Gelman-Rubin criterion \cite{ref66} to guarantee the convergence of our MCMC chains.  We have analysed the samples using the GetDist Python module \cite{ref67}. In the last row of Table \ref{tab2}, for the model comparison, we calculate the relative log-Bayesian evidence  ($\ln B_{ij}$) using the publicly accessible \texttt{MCEvidence} package ~\cite{ref68,ref69} to approximate the Bayesian evidence of extended $\Lambda_{\rm s}$CDM model relative to the extended $\Lambda$CDM model.  we make use of the updated Jeffrey's scale introduced by Trotta ~\cite{ref70}. We classify the evidence's strength as follows: it is considered inconclusive when $0 \leq | \ln B_{ij}|  < 1$, weak if $1 \leq | \ln B_{ij}|  < 2.5$, moderate if $2.5 \leq | \ln B_{ij}|  < 5$, strong if $5 \leq | \ln B_{ij}|  < 10$, and very strong if $| \ln B_{ij} | \geq 10$.


\section{Results and discussion}\label{sec:results}

 In Table \ref{tab2}, we present marginalized constraints at  68\% CL on the baseline parameters $(\omega_b, \omega_c, \theta_s, A_s, n_s, \tau_{reio}, z_{\dagger})$ and key derived parameters $(H_0, M_B, \Omega_m, \sigma_8, S_8)$ for the $\Lambda_s$CDM and $\Lambda$CDM models, based on different combinations of datasets such as Pk18, Pk18+DESI BAO, and Pk18+DESI BAO+PP\&SH0ES. At first, we discuss the impact of free parameter $z_{\dagger}$ on other cosmological parameters. The  Pk18 analysis reveals a flat one-dimensional marginalized distribution for $z_\dagger$ within the range $[1,3]$, so it cannot be constrained. When joint analysis of DESI BAO with Pk18, there are only lower bound exist, but a clear peak shape for $z_\dagger$  does not appear. A similar trend occurs  when joint analysis PP\&SH0ES  with the Pk18+DESI BAO datasets in Fig.\ref{fig1}. \\


\begin{figure}[hbt!]
	\centering
	\includegraphics[width=0.74\linewidth]{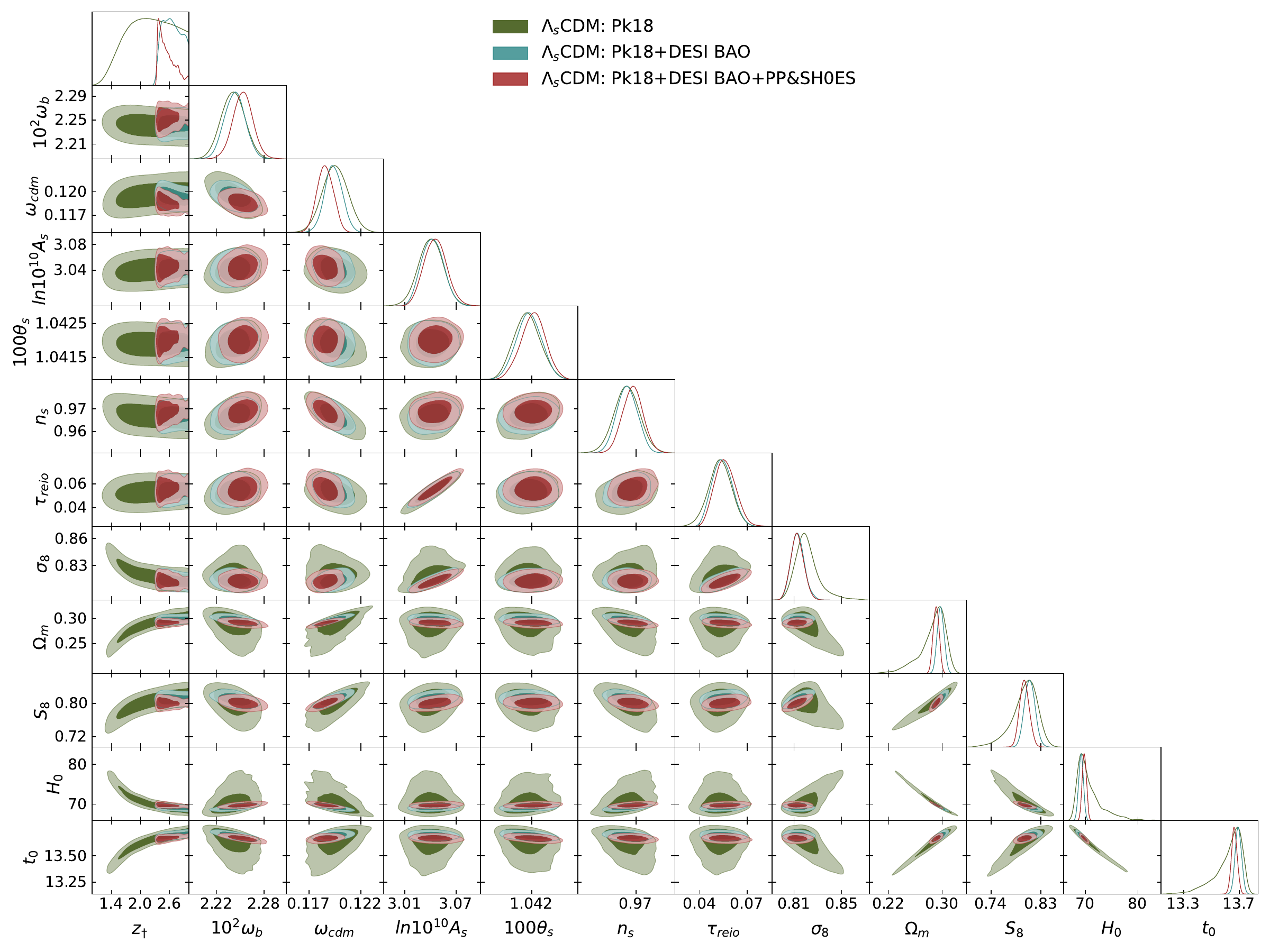}
	\caption{Marginalized one- and two-dimensional distributions (68$\%$ and 95$\%$ CLs)  of the $\Lambda_s$CDM model parameters for different datasets combinations: Pk18,
		Pk18+ DESI BAO, and Pk18+ DESI BAO+PP$\&$SH0ES.}
	\label{fig1}
\end{figure}

Now, we assess $H_0$ constraints obtained by combining various data sets with DESI in  $\Lambda_s$CDM and $\Lambda$CDM models. When using {Pk18}-only data, the value of $H_0$ is constrained to   $  {70.77}^{+0.79}_{ -2.7}$ $\text{km} \text{s}^{-1} \text{Mpc}^{-1}$ in the $\Lambda_s$CDM model and $ 67.39 \pm 0.55 $ $\text{km} \text{s}^{-1} \text{Mpc}^{-1}$ in the $\Lambda$CDM, both at a 68\% C.L.  From Pk18+DESI BAO (Pk18+DESI BAO+PP\&SH0ES) combinations yield $H_0$ values of $69.17\pm 0.44$ ($69.80\pm 0.40$) $\text{km} \text{s}^{-1} \text{Mpc}^{-1}$ under $\Lambda_s$CDM, as shown in Table \ref{tab2}, while $H_0$ is constrained to $68.31\pm 0.38$ ($68.92\pm 0.36$) $\text{km} \text{s}^{-1} \text{Mpc}^{-1}$ for the $\Lambda$CDM model. Overall, $\Lambda_s$CDM provides consistently {higher mean $H_0$ values across the data combinations than $\Lambda$CDM}. Quantifying the $H_0$ tension with SH0ES data $(H_{0}^{R22}=73.04\pm1.04$ $ \text{km} \text{s}^{-1} \text{Mpc}^{-1})$. From  Table \ref{tab3}, the $\Lambda_s$CDM and $\Lambda$CDM models show $3.4\sigma$ and $4.3\sigma$ $H_0$ tensions, respectively, with Pk18+BAO DESI data. This indicates that the $\Lambda_s$CDM model reduces the $H_0$ tension approximately by $0.9\sigma$. With the inclusion of Pk18+DESI BAO + PPSH0ES data, a $2.9\sigma$ tension is observed in the $\Lambda_s$CDM model, compared to $3.7\sigma$ for $\Lambda$CDM; in contrast, the overall tension is reduced by approximately $0.8\sigma$. Furthermore, we quantify the $H_0$ tension through TRGB analysis ($H_0^{\text{TRGB}} = 68.80 \pm 0.8$ $\text{km} \text{s}^{-1} \text{Mpc}^{-1}$). We also notice that in Table \ref{tab3},  the $H_0$ tension resolves within the $\Lambda_s$CDM framework compared to the $\Lambda$CDM model across all considered datasets.To constrain the $\Lambda_s$CDM model parameters, we also perform marginalized one qnd two dimentional (68$\%$ and 95$\%$ CLS) of the $\Lambda$ and model parameters for different data set combinations PK 18,PK18+DESI ~BAO and PK18+DESI ~BAO+PP$\&$ SHOES (see Fig.2).In our analysis  we observe that  the Fig.\ref{fig3} provides a comparative visualization of how different combinations of cosmological data sets and models ($\Lambda_{s}$CDM vs. $\Lambda$CDM) constrain the matter density and Hubble constant. It demonstrates the model-dependence of $H_{0}$ inference and highlights the persistent tension between Pk18-inferred and locally measured Hubble values.

\begin{figure}[hbt!]
	\centering
	\includegraphics[width=0.8\linewidth]{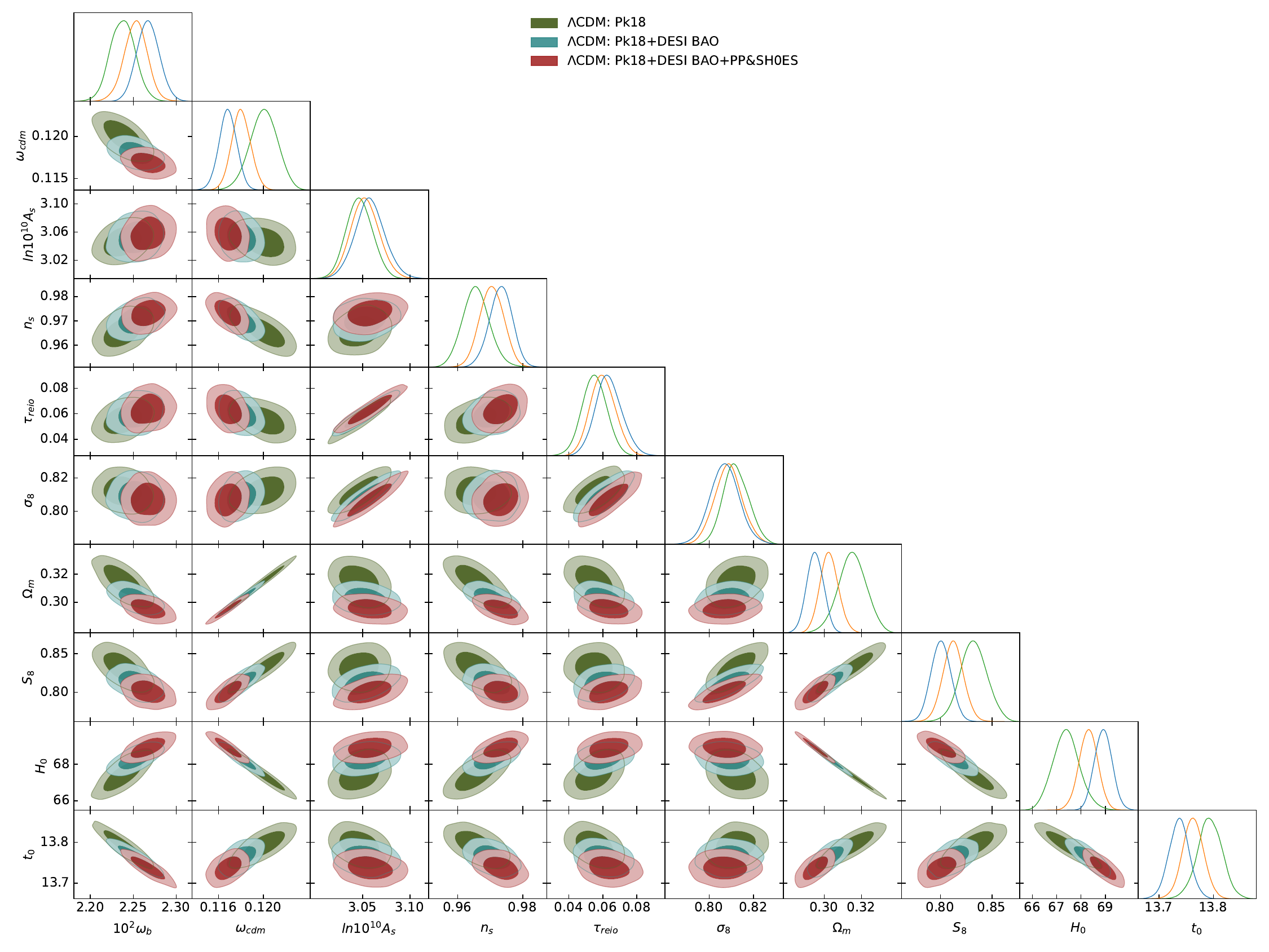}
	\caption{Marginalized one- and two-dimensional distributions (68\% and 95\% CLs)  of the $\Lambda$CDM model parameters for different datasets combinations: Pk18,
		Pk18+ DESI BAO, and Pk18+ DESI BAO+PP\&SH0ES.}
	\label{fig2}
\end{figure}

The left panel of Fig.\ref{fig3} is likely a contour plot or a comparison showing the constraints on $H_{0}$ and $\Omega_{m}$ from different models and datasets. The $\Lambda$CDM model is the standard cosmological model, while $\Lambda_s$CDM  might be a modified version. The data combinations affect the inferred values of $H_{0}$ and $\Omega_{m}$. For instance, including SH0ES data (which finds a higher $H_{0}$) might shift the $H_{0}$ values upwards compared to when only Pk18 and DESI BAO data are used. 
However, the figure shows , the $H_0$ values go from $71$ down to $67$, which might suggest that when adding more datasets (like PP$\&$SHOES), the $H_{0}$ constraint becomes lower? 
The key message is likely the tension in $H_0$ values between different models and datasets, showing how adding more data affects the parameter estimates.\\

\begin{figure*}[hbt!]
    \centering
    \includegraphics[width=0.4\linewidth]{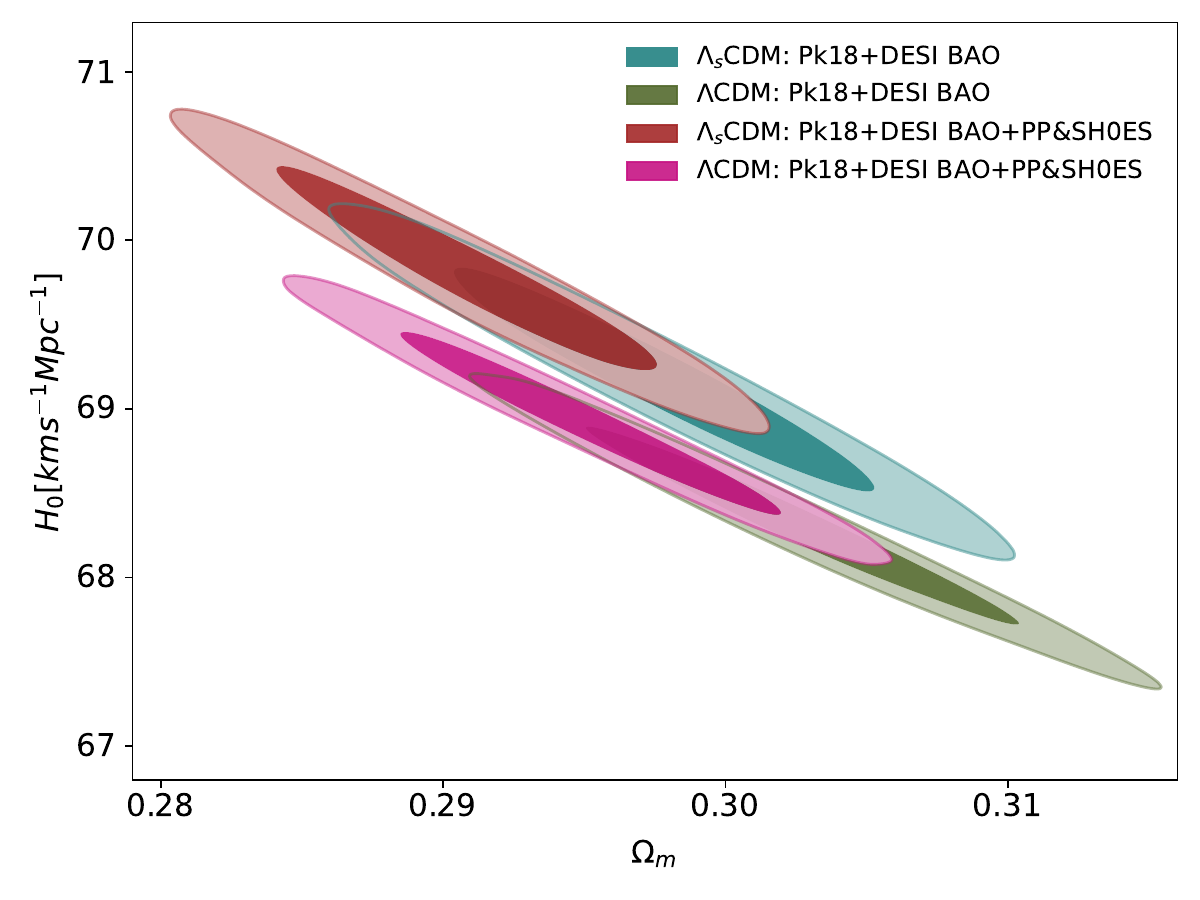}
    \includegraphics[width=0.4\linewidth]{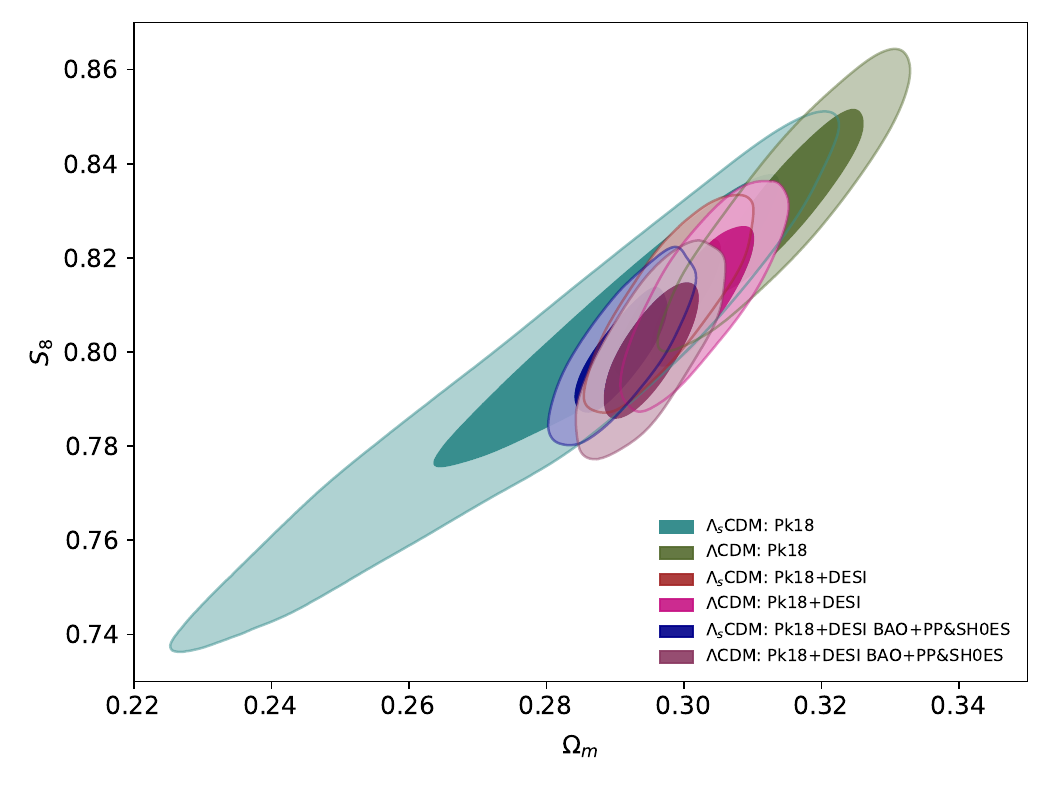}
     \includegraphics[width=0.4\linewidth]{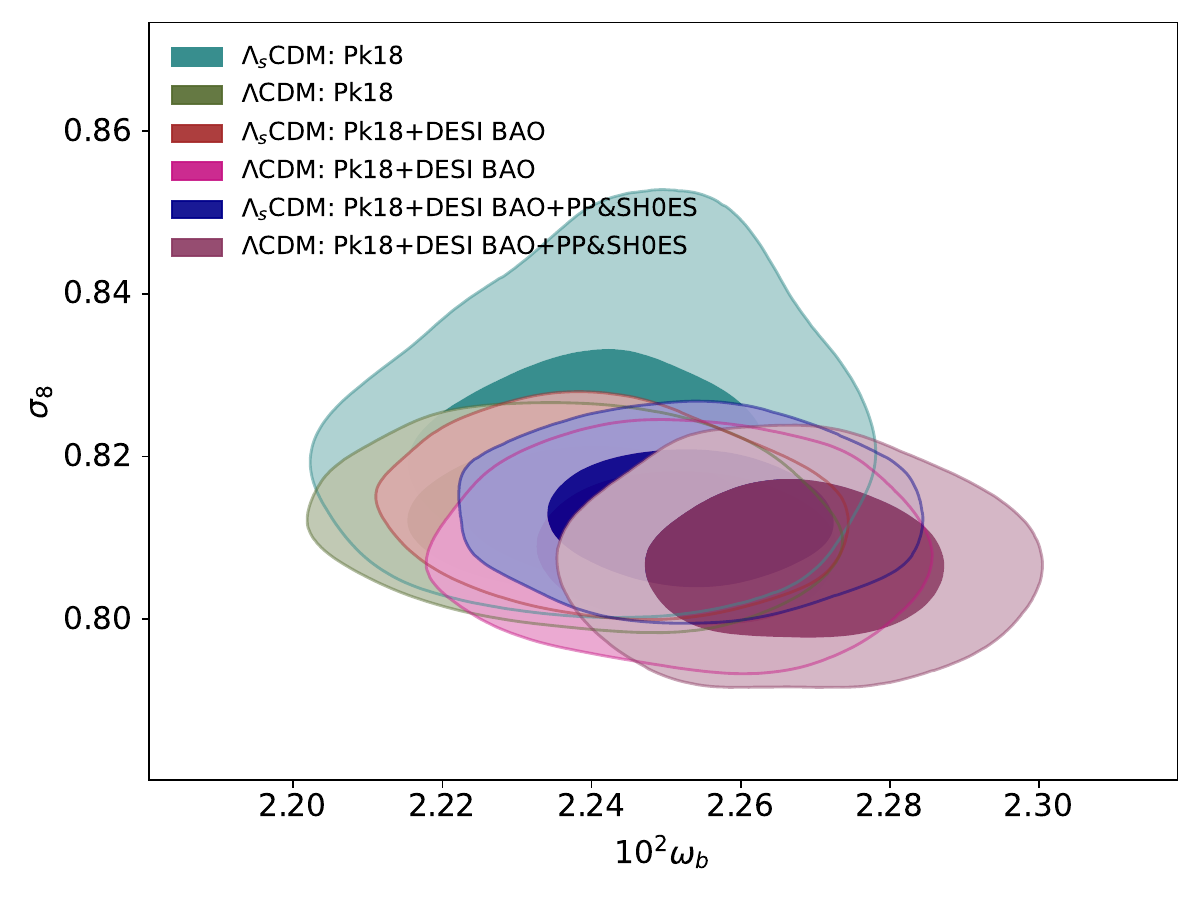}
    \caption{The left side 2D contour plots in the $H_0$-$\Omega_{\rm m}$ plane, and the right side 2D contour plots in the $S_8$-$\Omega_{\rm m}$ plane, and the middle bottem 2D contour plots in the $sigma_8$-$10^2\omega_{\rm b}$ plane  shown at 68$\%$ and 95$\%$ CL for the $\Lambda_{\rm s}$CDM and $\Lambda$CDM models with Pk18, Pk18+DESI BAO, and Pk18+DESI BAO+PP$\&$SH0ES.}
    \label{fig3}
\end{figure*}
\begin{table*}[hbt!]
	\renewcommand{\arraystretch}{1.3} 
	\caption{Marginalized constraints (mean values with 68\% CL) on the free and some derived parameters of the $\Lambda_{\rm s}$CDM and  $\Lambda$CDM models for different dataset combinations. The relative log-Bayesian evidence, $\ln \mathcal{B}_{ij} = \ln \mathcal{Z}_{\Lambda \rm CDM} - \ln \mathcal{Z}_{\Lambda_{\rm s} \rm CDM}$ is also displayed for each study in the last row; a negative value indicates that the $\Lambda_{\rm s}$CDM model is preferred over the $\Lambda$CDM scenario.}
	\label{tab2}
	
	\centering
	\scalebox{0.95}{
		\begin{tabular}{lccc}
			\toprule
			\textbf{Dataset}\;\;\;\;\;\;\;\; & \;\;\;\;\;\;\;\;\textbf{Pk18}\;\;\;\;\;\;\;\; &\;\;\;\;\;\;\;\; \textbf{Pk18+DESI BAO}\;\;\;\;\;\;& \textbf{Pk18+DESI BAO+PP\&SH0ES} \;\; \\
			\hline
			\hline
			
			\textbf{Model} & $\bm{\Lambda}_{\rm s}$CDM & $\bm{\Lambda}_{\rm s}$CDM & $\bm{\Lambda}_{\rm s}$CDM  \\
			& \textcolor{blue}{$\bm{\Lambda}$CDM} & \textcolor{blue}{$\bm{\Lambda}$CDM} & \textcolor{blue}{$\bm{\Lambda}$CDM}  \\
			
			\hline
			\hline
			
			$10^2\omega_b$ & $2.241 \pm 0.015$ & $2.243 \pm 0.013$ & $2.253 \pm 0.013$ \\
			& \textcolor{blue}{$2.238 \pm 0.014$} & \textcolor{blue}{$2.253 \pm 0.013$} & \textcolor{blue}{$2.267 \pm 0.013$} \\
			
			$\omega_{\rm cdm}$ & $0.1195 \pm 0.0012$ & $0.1194 \pm 0.0009$ & $0.1186 \pm 0.0008$ \\
			& \textcolor{blue}{$0.1200 \pm 0.0012$} & \textcolor{blue}{$0.1180 \pm 0.0008$} & \textcolor{blue}{$0.1168 \pm 0.0008$} \\
			
			$100\theta_s$ & $1.04189 \pm 0.00029$ & $1.04193 \pm 0.00028$ & $1.04203 \pm 0.00028$ \\
			& \textcolor{blue}{$1.04190^{+0.00027}_{-0.00031}$} & \textcolor{blue}{$1.04209 \pm 0.00027$} & \textcolor{blue}{$1.04224 \pm 0.00029$} \\
			
			$\ln(10^{10}A_s)$ & $3.040 \pm 0.014$ & $3.042 \pm 0.013$ & $3.046 \pm 0.013$ \\
			& \textcolor{blue}{$3.046 \pm 0.014$} & \textcolor{blue}{$3.052 \pm 0.015$} & \textcolor{blue}{$3.058 \pm 0.016$} \\
			
			$n_s$ & $0.9669 \pm 0.0043$ & $0.9670 \pm 0.0034$ & $0.9689 \pm 0.0034$ \\
			& \textcolor{blue}{$0.9657 \pm 0.0041$} & \textcolor{blue}{$0.9705 \pm 0.0036$} & \textcolor{blue}{$0.9733 \pm 0.0035$} \\
			
			$\tau_{\rm reio}$ & $0.0528 \pm 0.0073$ & $0.0543 \pm 0.0064$ & $0.0561^{+0.0062}_{-0.0071}$ \\
			& \textcolor{blue}{$0.0550 \pm 0.0072$} & \textcolor{blue}{$0.0600 \pm 0.0073$} & \textcolor{blue}{$0.0635^{+0.0072}_{-0.0081}$} \\
			
			$z_\dagger$ & unconstrained & $> 2.37$ (95\% CL) & $> 2.33$ (95\% CL) \\
			& \textcolor{blue}{$-$} & \textcolor{blue}{$-$} & \textcolor{blue}{$-$} \\
			
			\midrule
			
			$H_0$ [km/s/Mpc] & $70.77^{+0.79}_{-2.70}$ & $69.17 \pm 0.44$ & $69.80 \pm 0.40$ \\
			& \textcolor{blue}{$67.39 \pm 0.55$} & \textcolor{blue}{$68.31 \pm 0.38$} & \textcolor{blue}{$68.92 \pm 0.36$} \\
			
			$M_B$ [mag] & $-$ & $-$ & $-19.370 \pm 0.011$ \\
			& \textcolor{blue}{$-$} & \textcolor{blue}{$-$} & \textcolor{blue}{$-19.396 \pm 0.010$} \\
			
			$\Omega_m$ & $0.2860^{+0.0230}_{-0.0099}$ & $0.2977 \pm 0.0050$ & $0.2910 \pm 0.0044$ \\
			& \textcolor{blue}{$0.3151 \pm 0.0075$} & \textcolor{blue}{$0.3027 \pm 0.0049$} & \textcolor{blue}{$0.2951 \pm 0.0045$} \\
			
			$\sigma_8$ & $0.8210^{+0.0064}_{-0.0110}$ & $0.8131^{+0.0052}_{-0.0058}$ & $0.8127 \pm 0.0055$ \\
			& \textcolor{blue}{$0.8121^{+0.0055}_{-0.0061}$} & \textcolor{blue}{$0.8087 \pm 0.0062$} & \textcolor{blue}{$0.8070 \pm 0.0066$} \\
			
			$S_8$ & $0.801^{+0.026}_{-0.016}$ & $0.810 \pm 0.010$ & $0.800^{+0.008}_{-0.009}$ \\
			& \textcolor{blue}{$0.832 \pm 0.013$} & \textcolor{blue}{$0.812 \pm 0.010$} & \textcolor{blue}{$0.800 \pm 0.010$} \\
			
			$t_0$ [Gyr] & $13.62^{+0.12}_{-0.04}$ & $13.69 \pm 0.02$ & $13.67 \pm 0.02$ \\
			& \textcolor{blue}{$13.79 \pm 0.02$} & \textcolor{blue}{$13.76 \pm 0.02$} & \textcolor{blue}{$13.74 \pm 0.02$} \\
			
			\midrule
			
			$\chi^2_{\rm min}$ & $1389.03$ & $2793.72$ & $4105.66$ \\
			& \textcolor{blue}{$1390.26$} & \textcolor{blue}{$2797.92$} & \textcolor{blue}{$4119.32$} \\
			
			$\ln \mathcal{Z}$ & $-1423.17$ & $-1431.92$ & $-2088.73$ \\
			& \textcolor{blue}{$-1424.45$} & \textcolor{blue}{$-1433.99$} & \textcolor{blue}{$-2095.06$} \\
			
			$\ln \mathcal{B}_{ij}$ & $-1.28$ & $-2.07$ & $-6.33$ \\

		\hline
		\hline
		\end{tabular}
	}
	
\end{table*}

In the right panel of Fig.\ref{fig3} the contours show how different datasets and model assumptions constrain the parameters $\Omega_{m0}$ and $S_{8}$. The overlap of contours indicates consistency between datasets.
The standard $\Lambda$ Cold Dark Matter ($\Lambda$CDM) cosmological model offers an excellent fit to current observational data. However, notable and statistically significant tensions have arisen in the estimation of cosmological parameters when comparing results from the Planck satellite—focused on measuring anisotropies in the Cosmic Microwave Background (CMB)-with those from various low-redshift observational probes. Beyond the well-known discrepancy in the Hubble constant $H_{0}$, Planck data also show tension with weak lensing observations and galaxy redshift surveys concerning the matter density parameter $\Omega_{m}$and the amplitude or growth rate of cosmic structures (quantified by $\sigma_8$
and $f_{\sigma_8}$). While these discrepancies might stem from systematic uncertainties, they also motivate the exploration of potential new physics beyond the standard model.
The graph likely shows confidence contours or data points for each model and datasets combination in the $S_8$ vs $\Omega_m$ plane. The different datasets (Pk18, Pk18+DESI BAO,Pk18+DESI BAO+PP\&SH0ES) would show how adding more data affects the constraints on these parameters. The $\Lambda_{s}$CDM model might have an additional parameter compared to standard $\Lambda$CDM, leading to different allowed regions.

\begin{figure}[hbt!]
	\centering
	\includegraphics[width=0.75\linewidth]{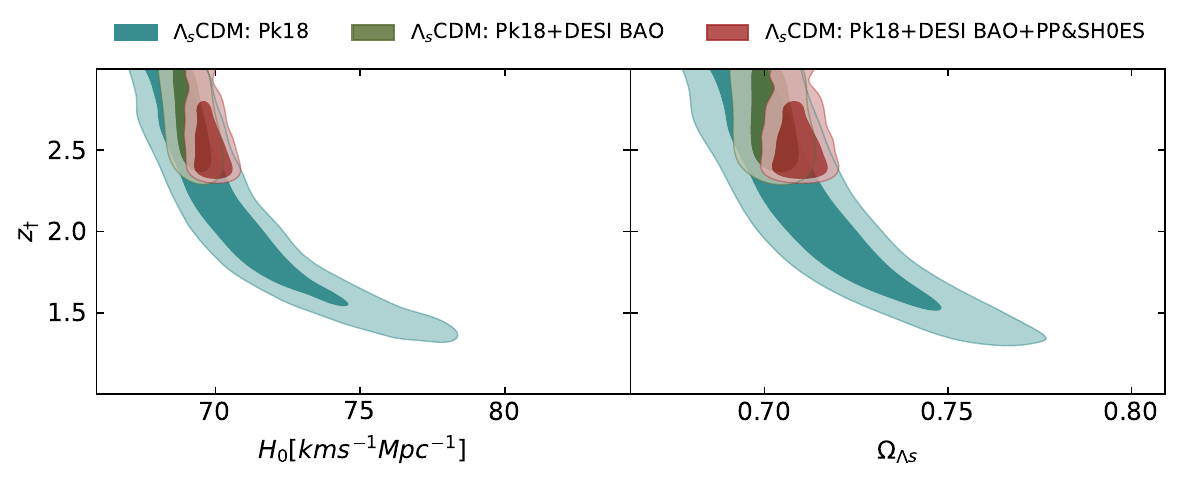}
	\caption{The left side 2D contour plots in the $z_{\dagger}$-$H_0$ plane, and the right side 2D contour plots in the $z_{\dagger}$-$\Omega_{\Lambda_{\rm s}}$ plane,shown at 68$\%$ and 95$\%$ CL for the $\Lambda_{\rm s}$CDM model with Pk18, Pk18+DESI BAO, and Pk18+DESI BAO+PP$\&$SH0ES.}
	\label{fig4}
\end{figure}
These tensions are often visualized in the $\sigma_8-\omega_{b}$ parameter space (see Fig.\ref{fig3} middle) and are commonly encapsulated in the parameter $S_{8}=\sigma_8\sqrt{\Omega_{m}/0.3}$, which aligns with the primary degeneracy direction in weak lensing data\cite{ref71}. The figure likely demonstrates how $\Lambda_{s}$CDM responds to cosmological constraints compared to $\Lambda$CDM, with a focus on resolving tensions like $H_{0}$ or $\omega_{b}$. The specific trends (e.g., whether $\Lambda_{s}$CDM improves fits) depend on the exact parameter shifts shown in the figure. If $\Lambda_{s}$CDM's $H_{0}$ aligns better with local measurements without violating Pk18 constraints, it could support dynamical dark energy or new physics beyond $\Lambda$CDM. The graph compares cosmological parameters $\sigma_8$(amplitude of matter fluctuations) and $\omega_{b}$ (baryon density) for two models, $\Lambda_{s}$CDM , using different observational datasets.So the graph is likely showing constraints on $\sigma_8$ and $\omega_{b}$ for different  combinations of datasets. Each model's constraints are plotted with different datasets, showing how adding more data (like DESI BAO and PP$\&$SHOES) affects the parameter estimates. The addition of datasets tightening the constraints would indicate reduced uncertainties when combining data.The baryon density $\omega_{b}$ is a parameter that affects Big Bang Nucleosynthesis and the Pk18, so its constraints are important for consistency across different observations.

\begin{table*}[hbt!]
\centering
\renewcommand{\arraystretch}{1.8} 
\setlength{\tabcolsep}{10pt} 
\caption{ A quantitative comparison between the key cosmological parameters of the $\Lambda_{\rm s}$CDM/$\Lambda$CDM models and the theoretical/direct measurements,
viz., $H_{0}^{\text{R21}} = 73.04 \pm 1.04$ km s$^{-1}$ Mpc$^{-1}$  and $H_{0}^{\text{TRGB}} = 69.8 \pm 0.8$ km s$^{-1}$ Mpc$^{-1}$;
$M_B = -19.244 \pm 0.037$ mag (SH0ES);
$S_8 = 0.834\pm0.016$ (Planck2018); $t_0 = 13.50 \pm 0.15$ Gyr (systematic uncertainties are not included);
$10^2\omega_{b}^{\text{LUNA}} = 2.233 \pm 0.036$  and
$10^2\omega_{b}^{\text{PCUV21}} = 2.195 \pm 0.022$ to assess the level of agreement.  }
\label{tab:comparison}

\scalebox{0.80}{

\begin{tabular}{l|c|cccccc}
\hline
\hline
\multirow{2}{*}{\textbf{Parameter}} & \multirow{2}{*}{\textbf{Observations}} & \multicolumn{2}{c}{\textbf{Pk18}} & \multicolumn{2}{c}{\textbf{Pk18+DESI BAO}} & \multicolumn{2}{c}{\textbf{Pk18+DESI BAO+PP\&SH0ES}} \\
\cmidrule(lr){3-4} \cmidrule(lr){5-6} \cmidrule(lr){7-8}
& & {\boldmath$\Lambda_s$}\textbf{CDM} & {\boldmath$\Lambda$}\textbf{CDM} & {\boldmath$\Lambda_s$}\textbf{CDM} & {\boldmath$\Lambda$}\textbf{CDM} & {\boldmath$\Lambda_s$}\textbf{CDM} & {\boldmath$\Lambda$}\textbf{CDM} \\
\hline
\hline
 
\multirow{2}{*}{\boldmath$H_0 {\rm[km/s/Mpc]}$}  & 
{\rm R21} & $1.1\sigma$ & $4.8\sigma$ & $3.4\sigma$ & $4.3\sigma$ & $2.9\sigma$ & $3.8\sigma$ \\
& {\rm TRGB} & $0.5\sigma$ & $2.5\sigma$ & $0.7\sigma$ & $1.7\sigma$ & $0.0\sigma$ & $1.0\sigma$  \\
\hline

\multirow{1}{*}{\boldmath$M_B[\rm mag]$} & {\rm SH0ES} & $-$ & $-$ & $-$ & $-$ & $3.2\sigma$ & $4.0\sigma$  \\

\multirow{1}{*}{\boldmath$S_8$} & {\rm \textcolor{red}{Planck2018}}& \textcolor{red}{$1.3\sigma$} & \textcolor{red}{$0.1\sigma$} & \textcolor{red}{$1.3\sigma$} & \textcolor{red}{$1.2\sigma$} & \textcolor{red}{$1.9\sigma$} & \textcolor{red}{$1.8\sigma$} \\

\multirow{1}{*}{\boldmath$t_0[\rm Gyr]$} & Direct & $0.5\sigma$ & $1.9\sigma$ & $1.3\sigma$ & $1.7\sigma$ & $1.1\sigma$ & $  1.6\sigma$ \\
\hline 
\multirow{2}{*}{\boldmath$\omega_{b}$ } & PCUV21 & $1.7\sigma$ & $1.6\sigma$ & $1.8\sigma$ & $2.2\sigma$ & $2.2\sigma$ & $2.8\sigma$ \\
& LUNA &$0.2\sigma$ & $0.1\sigma$ & $0.2\sigma$ & $0.5\sigma$ & $0.5\sigma$ & $0.8\sigma$ \\
\hline
\hline
\end{tabular}
}
 \label{tab3}
\end{table*}

In left pannel of  Fig.\ref{fig4}, we observe that the free parameter $ z_{\dagger} $ exhibits a positive correlation with $ H_0 $ in the extended model. As $ z_{\dagger} $ increases, the redshift at which key transitions—such as the onset of cosmic acceleration—occur also shifts to earlier times, potentially indicating a tighter and faster expansion phase. These models encompass even the largest model-independent measurements of $ H_0 $, reaching up to approximately $77 kms^{-1}Mpc^{-1} $ at 95\% confidence level in the PK18 data analysis. Due to this strong correlation, the constraints on $ z_{\dagger} $ directly influence the inferred values of $ H_0 $. However, when DESI BAO and Pantheon+SH0ES data are included in the analysis, this correlation weakens, and the allowed range for  $z_{\dagger} $ becomes more tightly constrained on the lower end. However, in the right panel of Fig.\ref{fig4}, a similar trend is observed between $ z_{\dagger} $ and $ \Omega_{\Lambda_{\rm s}}$ , suggesting that the timing of cosmic acceleration is also closely related to the present dark energy density in the extended model.\\

\begin{figure}[hbt!]
	\centering
	\includegraphics[width=0.75\linewidth]{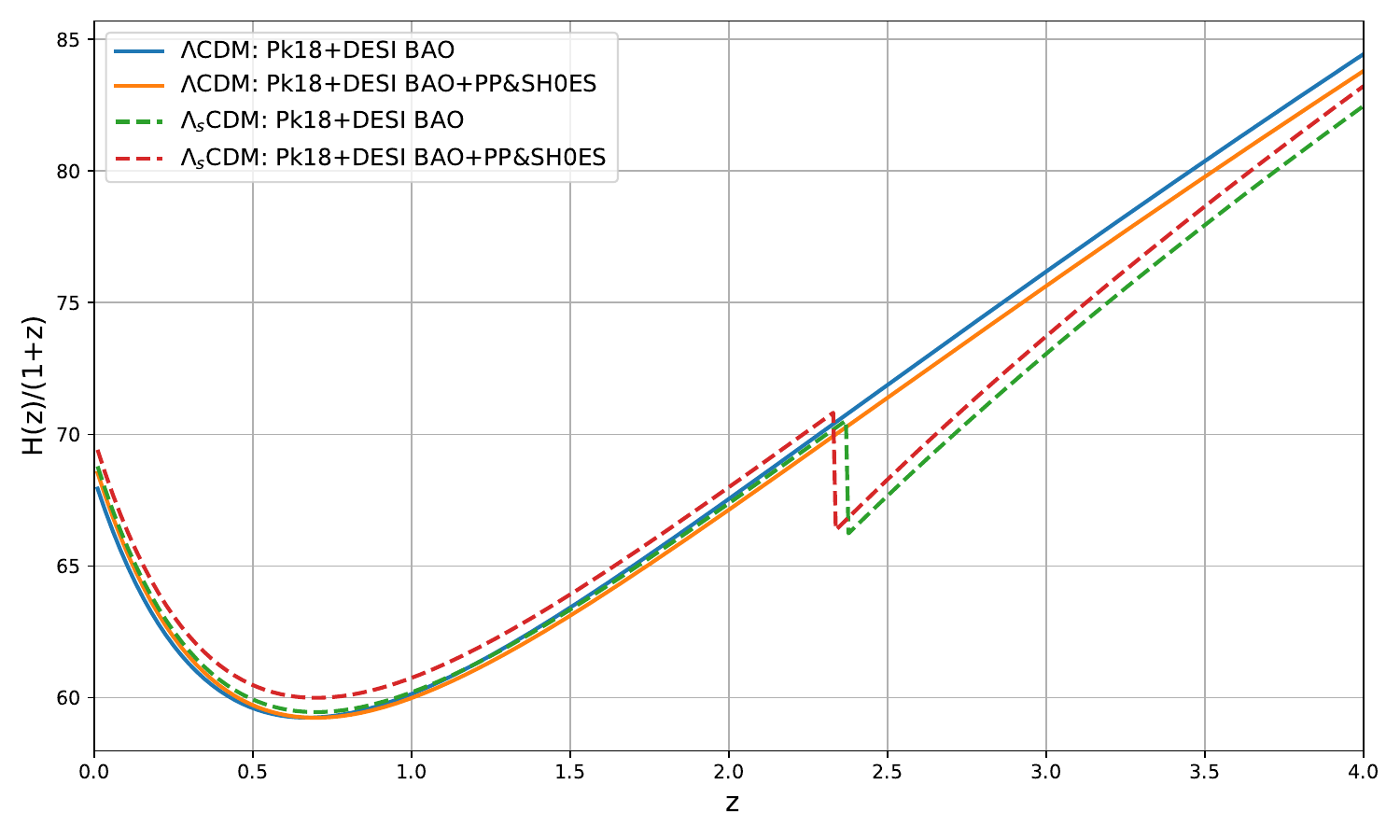}
	\caption{Statistical reconstruction of the rescaled expansion rate of the universe, $H(z)/(1 + z)$, at $1\sigma$ for the $\Lambda$CDM and $\Lambda_s$CDM models, based on the joint analysis of Pk18+DESI BAO, and Pk18+DESI BAO+PP$\&$SH0ES.}
	\label{fig5}
\end{figure}

One of the central $S_8$ tensions in modern cosmology is closely tied to the formation and growth of cosmic structures. The Planck 2018 cosmic microwave background (CMB) measurements, within the framework of the $\Lambda$CDM model, provide a best-fit value of $S_8 = 0.834 \pm 0.016$ \cite{ref72}, which exhibits a growing statistical tension—at the level of 2 to 3$\sigma$—when compared with measurements from weak gravitational lensing (WGL) surveys \cite{ref73}. For instance, the KiDS-1000 \cite{ref13} survey reported  $S_8 = 0.759^{+0.024}_{-0.021}$, indicating a $\sim$3$\sigma$ discrepancy with Planck. Similarly, the Hyper Suprime-Cam Year 3 analysis found $S_8 = 0.776^{+0.032}_{-0.033}$ \cite{ref74}, corresponding to a 2$\sigma$ tension, while the Dark Energy Survey Year 3 (DES-Y3) yielded $S_8 = 0.759^{+0.025}_{-0.023}$\cite{ref75}, reflecting a 2.3$\sigma$ tension. A combined analysis of cosmic shear data from DES-Y3 and KiDS-1000 provided an improved constraint of $S_8 = 0.790^{+0.018}_{-0.014}$ \cite{ref76}, reducing the tension with Planck to the 1.7$\sigma$ level. Most recently, the KiDS-Legacy survey reported $S_8 = 0.815^{+0.016}_{-0.021}$,\cite{ref77} which shows good agreement with Planck, with a reduced tension of just 0.73$\sigma$, suggesting a potential resolution of the longstanding $S_8$ discrepancy. In this context, our analysis shows that the $S_8$ parameter is constrained to a best-fit value of $0.801^{+0.026}_{-0.016}$ within the $\Lambda_s$CDM model using the Pk18 dataset. When extended to joint analyses with additional data, namely Pk18+DESI BAO and Pk18+DESI BAO+PP\&SH0ES, both the $\Lambda_s$CDM and $\Lambda$CDM models yield nearly identical mean values of $S_8$. Specifically, from the Pk18+DESI BAO dataset, the $S_8$ constraints are $0.810 \pm 0.010$ for $\Lambda_s$CDM and $0.812 \pm 0.010$ for $\Lambda$CDM. Similarly, from the full combination of Pk18+DESI BAO+PP\&SH0ES, the constraints are $0.800^{+0.008}{-0.009}$ for $\Lambda_s$CDM and $0.800 \pm 0.010$ for $\Lambda$CDM. These results indicate that, across all considered datasets, the $S_8$ values derived from both models remain consistent with the recent constraints from the KiDS Legacy survey, as well as the combined cosmic shear analysis of DES-Y3 and KiDS-1000, supporting the robustness and observational compatibility of our model. From Table \ref{tab3}, We observe that the $S_8$ tension is notably reduced in both models, reaching approximately $1.3\sigma$ when using the Pk18+DESI BAO dataset and around $1.8\sigma$ with the full combination of Pk18+DESI BAO+PP\&SH0ES. This reduction in $S_8$ tension highlights the consistency of our results with weak lensing measurements. We observe that in right panel Fig.\ref{fig3} shows the 68\% and 95\% C.L contours in the $S_8-\Omega_m$ plane, which strongly overlap contours as well as positive correlation in both models with Pk18+DESI BAO and Pk18+DESI BAO+PP\&SH0ES data.\\

 From Fig.\ref{fig5} provides a comparative analysis between the standard $\Lambda$CDM model and the sign-switching dark energy model $\Lambda_s$CDM by plotting ${H(z)}/(1+z)$ against $z$. It includes two dataset combinations: Pk18+DESI BAO and Pk18+DESI BAO+PP\&SH0ES. The aim is to examine how closely the two models align across different cosmic epochs. From this figure, we observe that at high redshifts ($z > 3.5$), both $\Lambda$CDM and $\Lambda_s$CDM models show nearly identical behavior, indicating that the sign-switch model remains consistent with the standard model in the early universe. This agreement supports the idea that any deviations from $\Lambda$CDM in the $\Lambda_s$CDM framework become relevant only at low redshifts, where dark energy starts to dominate. At low redshifts, especially when PP\&SH0ES and DESI BAO data are included, small differences emerge between the two models. The $\Lambda_s$CDM model slightly deviates from $\Lambda$CDM, particularly in the range $1 < z < 3$, which reflects the influence of the sign-switching behavior in the late-time evolution of dark energy. However, these differences remain within observational uncertainties, suggesting that the $\Lambda_s$CDM model provides a viable alternative while still agreeing well with current data.\\

Our results are presented for $t_0$ in Table \ref{tab2} and concordance between the $\Lambda_s$CDM and $\Lambda$CDM models listed in Table \ref{tab3}. We reveal that in the Pk18+DESI BAO dataset, the $\Lambda$CDM model has a $1.7\sigma$ tension with globular clusters (GCs), while the $\Lambda_s$CDM model reduces this to $1.3\sigma$, leading to a $0.4\sigma$ decrease in $t_0$ tension. Similarly, for Pk18+DESI BAO+PP\&SH0ES, the $\Lambda$CDM model shows a $1.6\sigma$ tension, while the $\Lambda_s$CDM model lowers it to $1.1\sigma$, further reducing $t_0$ tension by $0.5\sigma$. Overall, the $\Lambda_s$CDM model aligns better agreement of the universe’s age with  GCs  than the $\Lambda$CDM model  from all considered data sets.\\


     \begin{figure}[hbt!]
    \begin{flushleft}
    \hspace*{1.5cm}
    \includegraphics[width=0.8\linewidth]{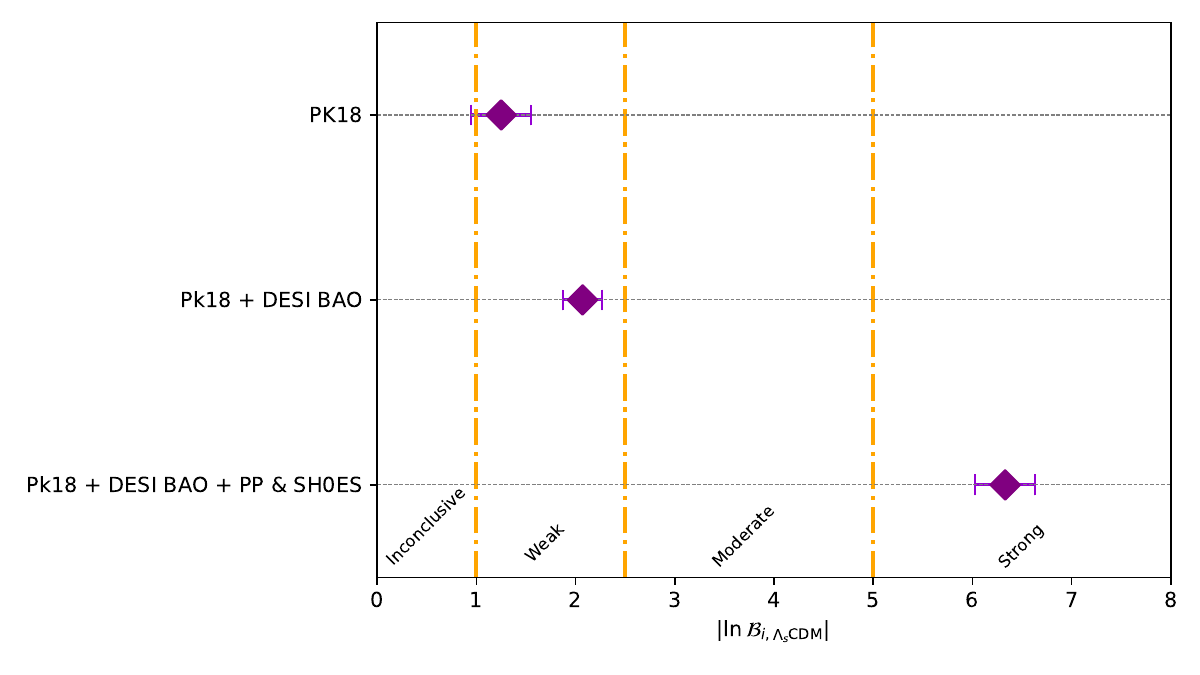}
\end{flushleft}
    
    \caption{Bayesian analysis of the  $\Lambda_s$CDM model in relation to the $\Lambda$CDM model.}
    \label{fig6}
\end{figure}

Finally, we assess which model is more effective by calculating the Bayesian evidence and applying the Jeffreys' scale for model comparison. The last three row in Table \ref{tab2}  listed that, the $\chi^{2}_{\rm min}$ and log-Bayesian evidence ($\ln \mathcal Z$) value for $\Lambda_s$CDM and $\Lambda$CDM models, as well as the Bayes’ factor  ($\ln \mathcal{B}_{ij} =| \ln \mathcal{Z}_{\Lambda \rm CDM} - \ln \mathcal{Z}_{\Lambda_{\rm s} \rm CDM}|$), which quantifies the difference in log-Bayesian evidence between $\Lambda$CDM model and the $\Lambda_s$CDM model. We notice in Table \ref{tab2} that between  $\Lambda_s$CDM and $\Lambda$CDM models perform inconclusive based on Bayesian evidence across Pk18 datasets. Also, we find weak Bayesian evidence  ($\ln \mathcal{B}_{ij}=2.07$) between $\Lambda_s$CDM and $\Lambda$CDM models from Pk18+DESI BAO datasets, but $\Lambda_s$CDM strong Bayesian statistical evidence ($\ln \mathcal{B}_{ij} = 6.33$) against $\Lambda$CDM from  Pk18+DESI BAO + PP\&SHOES data analysis.  The Fig.\ref{fig6} demonstrates that combining multiple datasets (Pk18, DESI BAO, PP\&SH0ES) significantly strengthens the statistical evidence for cosmological parameter constraints, resolving tensions or uncertainties present in individual datasets.This suggests the graph is showing the strength of evidence or statistical confidence for different combinations of datasets.The vertical bars with labels like ``PK18", ``PK18 + DESI BAO", etc., might indicate how combining more datasets increases the confidence level. For example, PK18 alone might have weak evidence, but adding DESI BAO and others moves the bar to moderate or strong.

\newpage
\section{Comparative Analysis of DESI BAO and DR2 Datasets }\label{sec:results}

In this comprehensive analysis, we investigate the implications of our proposed cosmological model by employing the recently released DESI BAO DR2 dataset \cite{ref78} label as DR2 in combination with the PP\&SH0ES observations. These high-precision datasets provide detailed insights into the universe’s expansion history and serve as powerful tools for constraining cosmological parameters. As shown in Fig.\ref{fig7}, the posterior distribution of the free parameter \( z_{\dagger} \) displays a prominent and well-defined peak. This parameter is tightly constrained to \( z_{\dagger} = 1.39^{+0.22}_{-0.31} \), with clearly established upper and lower bounds, indicating the model’s effectiveness in capturing key features of cosmic expansion. From the combined DR2 and PP\&SH0ES data, we also obtain a best-fit value of the Hubble constant \( H_0 = 73.05 \pm 0.95 \) km/s/Mpc, which is in excellent agreement with the SH0ES local measurements, thereby reinforcing the consistency of our model with nearby observational data. Additionally, Fig.\ref{fig7} reveals a slight negative correlation between \( H_0 \) and \( z_{\dagger} \), suggesting that higher values of the transition redshift correspond to slightly lower values of the Hubble constant. Our analysis further provides estimates of other key cosmological quantities: the current age of the universe is constrained to \( t_0 = 12.80 \pm 0.19 \) Gyr, and the matter density parameter is determined to be \( \Omega_{m} = 0.308^{+0.011}_{-0.0093} \), indicating that approximately 30\% of the total energy content of the universe consists of matter.\\
\begin{figure}[hbt!]
	
	\includegraphics[width=0.8\linewidth]{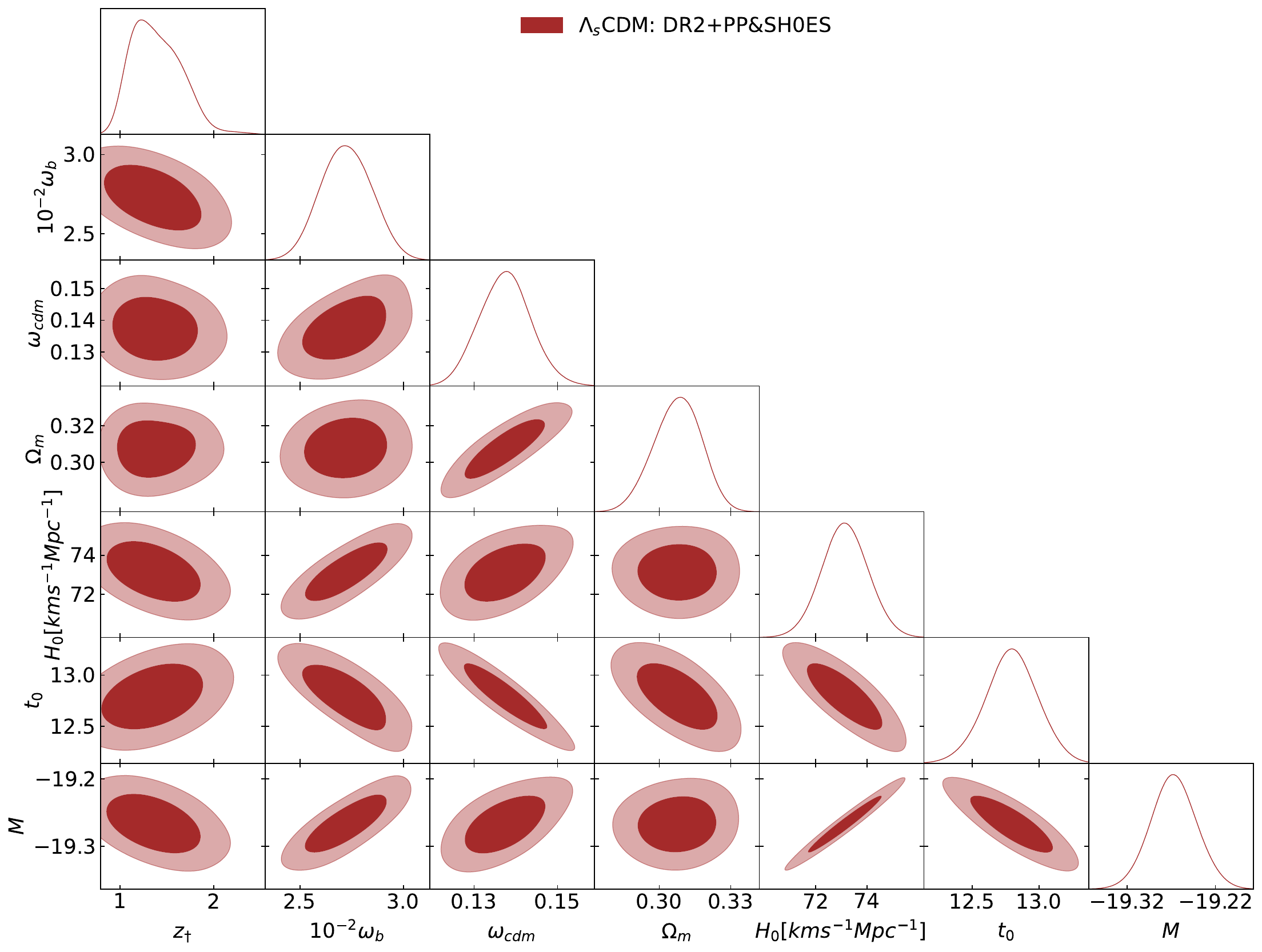}
	
	\caption{Marginalized one- and two-dimensional distributions (68\% and 95\% CLs)  of the $\Lambda_s$CDM model parameters for different datasets combinations: DR2+PP\&SH0ES.}
	\label{fig7}
\end{figure}

\begin{figure*}[tbh!]
	\centering
	\includegraphics[width=0.4\linewidth]{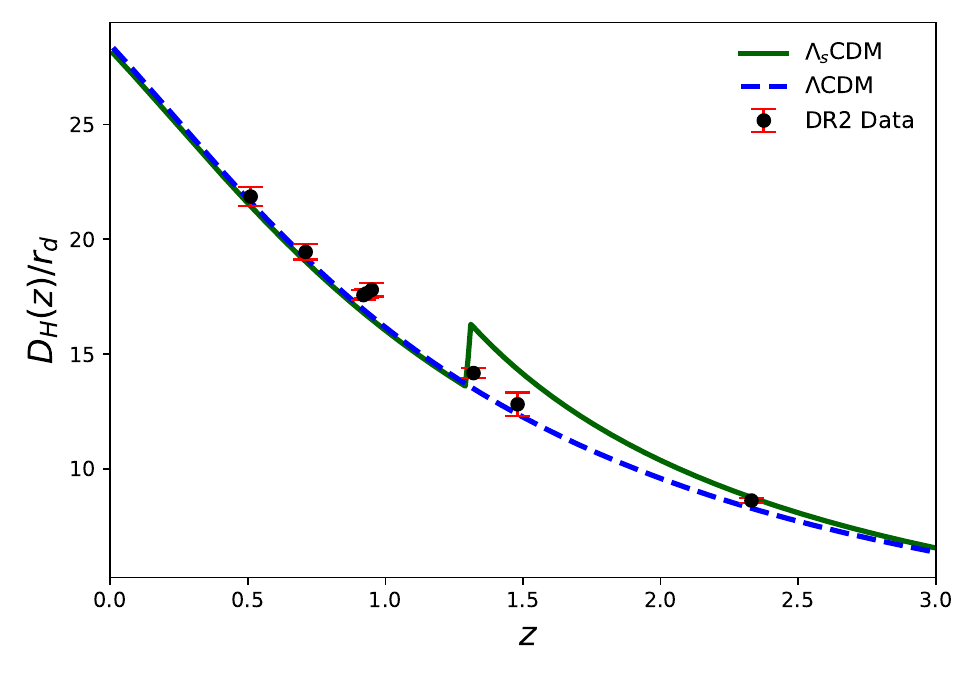}
	\includegraphics[width=0.4\linewidth]{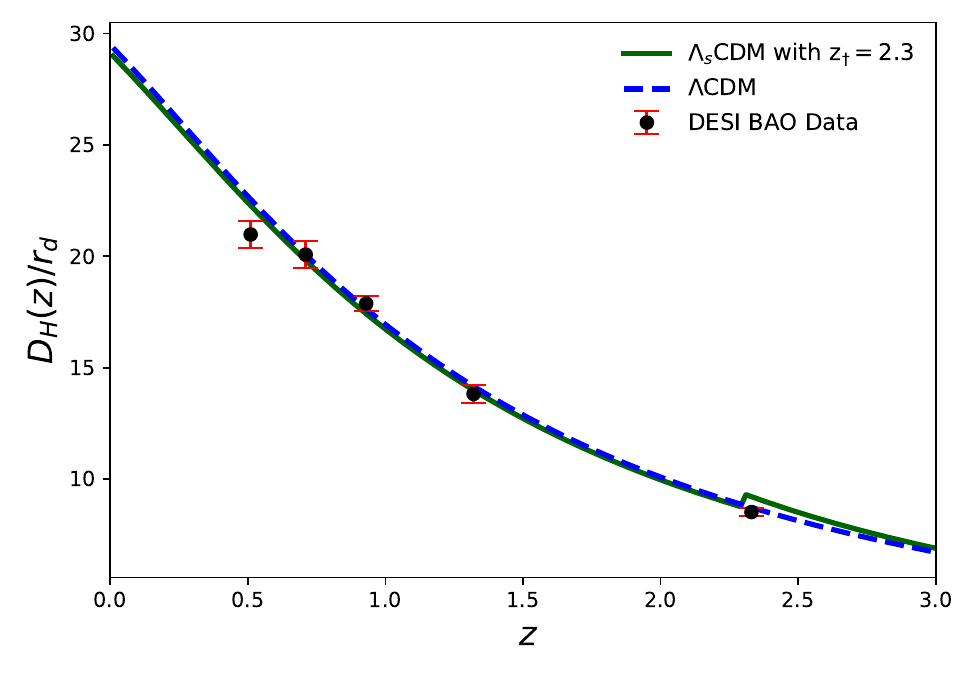}
	\includegraphics[width=0.4\linewidth]{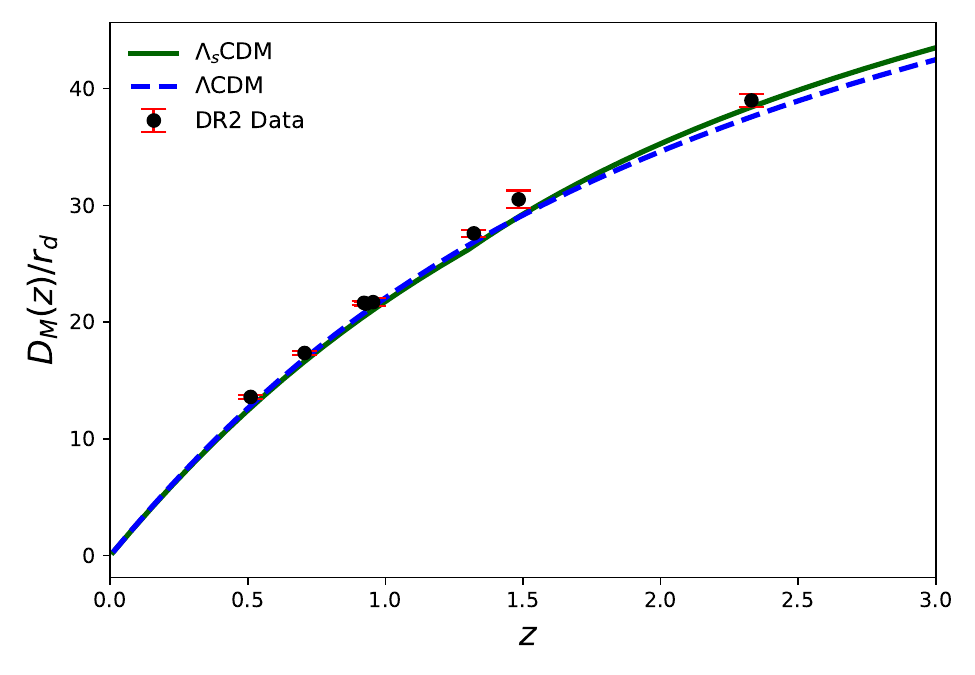}
	\includegraphics[width=0.4\linewidth]{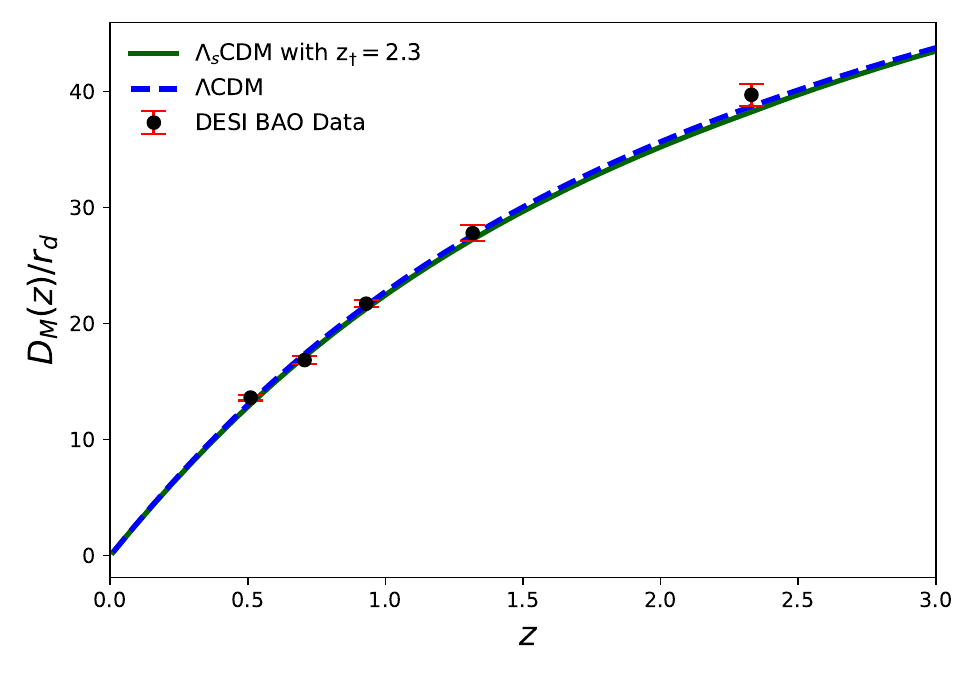}
	\caption{\rm The best-fit distance--redshift relations for the $\Lambda$CDM and $\Lambda_{\mathrm{s}}$CDM models are shown, derived from the analysis of the CMB+DESI  BAO+PP\&SH0ES  and DR2+PP\&SH0ES datasets. The predictions are presented along with their corresponding error bars.
	}
	\label{fig8}
\end{figure*}

In the top-left panel of Fig.\ref{fig8}, we present a comparative analysis of the quantity $D_H(z)/r_d$ as a function of redshift $z$. The dotted blue curve corresponds to the theoretical prediction from the standard $\Lambda$CDM model, while the dark green curve represents the prediction from our model, constrained using the combined DR2+PP\&SH0ES dataset. Observational measurements from the DR2 sample are indicated by black points with red error bars. Overall, the distance predictions from both models exhibit broad consistency, suggesting that our model remains compatible with the null hypothesis embodied by $\Lambda$CDM. Nevertheless, noticeable deviations between the theoretical curves are observed, which may indicate the presence of dynamical effects or interactions incorporated in our framework. In particular, the DR2 measurements of $D_H(z)/r_d$ at $z = 0.93$ and $z = 1.51$ exhibit tension with the predictions of our model. Furthermore, in the bottom-left panel of Fig.\ref{fig8}, the data point for $D_M(z)/r_d$ at $z = 1.51$ shows a clear deviation from both our model and the standard model, whereas the measurement at $z = 2.33$ demonstrates better agreement with our model compared to $\Lambda$CDM.

In the top-right panel of Fig.\ref{fig8}, we similarly compare $D_H(z)/r_d$ as a function of $z$, with predictions from the $\Lambda$CDM model (dotted blue) and our model constrained by the CMB+DESI BAO+PP\&SH0ES dataset (dark green). The DESI BAO measurements, depicted as black points with red error bars, generally good agreement with both models, with the exception of the data point at $z = 0.51$, which shows a notable deviation from both models predictions. In the bottom-right panel, the DESI BAO measurement of $D_M(z)/r_d$ at $z = 2.33$ also deviates significantly from the predictions of both models, indicating a potential feature in the data that warrants further investigation.

We conclude that the free parameter \( z_{\dagger} \) in our model exhibits varying degrees of constraint depending on the dataset employed. For the CMB+DESI BAO+PP\&SH0ES combination, the posterior distribution of \( z_{\dagger} \) displays a clear peak around \( z_{\dagger} = 2.3 \), but remains unconstrained from above, indicating a relatively weak constraint. In contrast, the DR2+PP\&SH0ES dataset provides a much tighter constraint, with a well-defined peak and clearly established bounds at \( z_{\dagger} = 1.39^{+0.22}_{-0.31} \). Furthermore, our model offers a more relaxed fit to the DR2 measurement of \( D_M(z)/r_d \) at \( z = 2.33 \), effectively alleviating the discrepancy observed with the standard \( \Lambda \)CDM model. This suggests that our model provides a better description of the high-redshift distance data in the DR2 combination.

\section{Conclusion}\label
In this work, we have investigated the recently proposed $\Lambda_{\rm s}$CDM model, characterized by a sign-switching cosmological constant, using the latest observational data from Pk18, DESI BAO, and PP\&SH0ES compilations. Our primary focus is to explore how the additional free parameter $z_{\dagger}$ influences other cosmological parameters and whether $\Lambda_{\rm s}$CDM offers a better fit to current data compared to the standard $\Lambda$CDM model.Our analysis shows that $z_{\dagger}$ remains largely unconstrained by the Pk18 data alone, while the addition of DESI BAO and PP\&SH0ES introduces a lower bound but does not lead to a clear detection. 
In the present analysis, the transition redshift  $z_{\dagger}$, which governs the sign-switching behavior of the cosmological constant $\Lambda_{\rm s}$CDM   remains largely unconstrained. This is due to the limited sensitivity of current observational data to such sharp transitions in the cosmic expansion rate. Future large-scale surveys such as Euclid, DESI, and the Roman Space Telescope are expected to improve this situation by providing higher-precision measurements across a broader redshift range. This suggests that while current datasets are starting to probe the new physics introduced by $\Lambda_{\rm s}$CDM, more precise data will be necessary to fully constrain the model.\\

In this paper, we have constrained a baseline and some derived parameters for $\Lambda_s$CDM, an extension of {$\Lambda_s$CDM},  models using different combinations of data sets, including Pk18, DESI BAO, and PP\&SHOES. Our analysis reveals that the Pk18 data do not constrain the free parameter $z_\dagger$. However, only a lower bound exists of $z_\dagger$ from Pk18+ DESI BAO and Planck+ DESI BAO +PPSH0ES data combination analysis. Interestingly, the DESI BAO data has no significant effect on $z_\dagger$, as the results for $z_\dagger$ remain the same with BAO data in the $\Lambda_s$CDM model \cite{ref14}. Besides, we have observed that the $\Lambda_s$CDM model estimates the higher values of the Hubble constant  $H_0 = 69.17 \pm 0.44  ( 69.80 \pm 0.40) km s^{-1} Mpc^{-1}$ from Pk18+DESI BAO (Pk18+DESI BAO+PP\&SH0ES) data analysis, respectively. These derived $H_0$ values both consider data are aligned with TRGB measurements but still exhibit tension with SH0ES $(H_{0}^{R22}).$ Further, We observe a slight impact of DESI BAO data on $H_0$ in the $\Lambda_s$CDM model when combined with Pk18 or Pk18 and PP\&SHOES data, leading to a slight increase in $H_0$ values; due to this, the $H_0$ tension are reduced by $0.8\sigma$ .\\
Notably, we find that both $\Lambda_{\rm s}$CDM and its fully predictive extension, $\Lambda_{\rm s}$VCDM, consistently achieve a lower $\chi^2$ compared to $\Lambda$CDM across different data combinations. This points toward a mild preference for models allowing for a dynamic evolution of the cosmological constant, especially considering the recently reported $3.9\sigma$ tension between preliminary DESI results and the $\Lambda$CDM model. However, given the current uncertainties and model dependencies, we cannot definitively claim that $\Lambda_{\rm s}$CDM, or any dynamical dark energy model, is favored over the standard cosmological constant.
Future DESI data releases, along with next-generation CMB and supernova surveys, will be crucial to testing the viability of $\Lambda_{\rm s}$CDM. In particular, improved constraints on BAO measurements at different redshifts and refined priors on dark energy parameters will help clarify whether a sign-switching cosmological constant offers a true resolution to the emerging tensions in cosmology.Thus, while $\Lambda_{\rm s}$CDM presents an intriguing and theoretically motivated alternative, more observational evidence is needed to assess
its role in the evolving picture of cosmic 
acceleration. 

\section*{Declaration of competing interest}
The authors declare that they have no known competing financial
interests or personal relationships that could have appeared to influence
the work reported in this paper.

\section*{Data availability}
No data was used for the research described in the article.

\section*{acknowledgments} 
The authors (A. Dixit \& A. Pradhan) are thankful to the Inter-University Centre for Astronomy and Astrophysics (IUCAA), Pune, India for providing support and facility under Visiting Associateship programs. M. Yadav is supported by a Junior Research Fellowship (CSIR/UGC Ref. No. 180010603050) from the University Grants Commission, Govt. of India. The authors thank the anonymous reviewers for their insightful remarks, which helped improve the manuscript in its current form.}

\end{document}